\documentclass[a4paper]{jpconf}
\usepackage{citesort}
\usepackage[dvips]{graphicx,color}
\usepackage{amsmath,amssymb}
\usepackage{bm}
\begin{document}
\title{Thermodynamic properties of quadrupolar states in the frustrated pyrochlore magnet Tb$_2$Ti$_2$O$_7$}

\author{H. Takatsu$^{1,2}$, T. Taniguchi$^2$, S. Kittaka$^3$, T. Sakakibara$^3$, and H. Kadowaki$^2$}
\address{$^1$Department of Energy and Hydrocarbon Chemistry, Graduate School of Engineering, Kyoto University, Kyoto 615-8510, Japan}
\address{$^2$Department of Physics, Tokyo Metropolitan University, Hachioji-shi, Tokyo 192-0397, Japan}
\address{$^3$Institute for Solid State Physics, University of Tokyo, Kashiwa 277-8581, Japan}

\begin{abstract}
The low-temperature thermodynamic properties of the frustrated pyrochlore 
Tb$_{2+x}$Ti$_{2-x}$O$_{7+y}$ have been studied  using the single crystal of $x=0.005$ 
sitting in a long range ordered phase in the $x$--$T$ phase diagram. 
We observed that the specific heat exhibits a minimum around 2~K and slightly increases on cooling,
similar to a Schottky-like anomaly for canonical spin ices.
A clear specific-heat peak observed at $T_{\rm c} = 0.53$~K is ascribable to the phase transition to a quadrupolar state,
which contributes to a relatively large change in entropy, $S \simeq 2.7$~J~K$^{-1}$mol$^{-1}$.
However, it is still smaller than $R\ln2$ for the ground state doublet of the Tb ions.
The entropy release persists to higher temperatures,
suggesting strong fluctuations associated with spin ice correlations above $T_{\rm c}$.
We discuss the field dependence of the entropy change for $H||[111]$ and $H||[001]$.
\end{abstract}

\section{Introduction}
Geometrically frustrated magnets have attracted much attention because of 
the realization of new type of electronic and 
magnetic phenomena with unconventional order parameters~\cite{C.Lacroix,GingrasRPP2014}. 
In particular, the pyrochlore-lattice magnet Tb$_{2+x}$Ti$_{2-x}$O$_{7+y}$,
a putative candidate of quantum spin liquid (QSL)~\cite{GardnerPRL1999,GardnerPRB2003},
shows unique properties including an unknown long range order (LRO) in the vicinity ($x_c = -0.0025 \leq x < 0.04 $) of 
the QSL state~\cite{TaniguchiPRB2013,M.WakitaJPCS2016}.
Indeed a clear specific-heat peak was observed at $T_{\rm c} \simeq 0.5$~K for the sample with $x = 0.005$, 
while no LRO associated with the large magnetic and/or structural phase transitions was confirmed~\cite{TaniguchiPRB2013}.
Although the only small Bragg peak with the order of 0.1~$\mu_{\rm B}$/Tb appears below $T_{\rm c}$,
it is too small to explain the corresponding entropy change in the specific heat.
It is thus apparently different from the magnetic dipole order inferred by earlier theories~\cite{GingrasPRB2000,KaoPRB2003}.
This mysterious, or hidden, order is an important subject for the study of 
the actual nature of the ground state of Tb$_2$Ti$_2$O$_7$~\cite{MolavianPRL2007,BonvillePRB2011,PetitEPJWC2015,GuittenyPRB2015,FritschPRB2013,FritschPRB2014,E.KermarrecPRB2015,FennellPRL2014,M.RuminyPRB2016-1,M.Ruminy2016}.

Recently, we have investigated the hidden order of Tb$_{2+x}$Ti$_{2-x}$O$_{7+y}$ using 
a single crystalline sample with $x = 0.005$ ($T_{\rm c} = 0.53$~K)
by means of neutron scattering, specific heat, magnetization measurements~\cite{KadowakiSPIN2015,H.TakatsuPRL2016}.
From the semi-quantitative analysis based on the theoretical model proposed by Onoda and Tanaka~\cite{S.Onoda2010PRL,S.Onoda2011PRB},
we have demonstrated that the ordered state originates from electric quadrupole moments inherent in the non-Kramers ion of Tb$^{3+}$.
It is remarkable that the estimated parameter set is located very close to the phase boundary between
the quadrupolar and U(1) QSL states. 
This result naturally explains the previous experimental result that the minute change in 
$x$ induces a phase transition between the QSL and LRO states~\cite{TaniguchiPRB2013,M.WakitaJPCS2016}.
These results also showed remarkable behaviors and possibilities for magnetic field, 
such as a two dimensional (2D) quadrupole order for $H||[111]$,
where the system behaves as decoupled 2D kagom\'{e} layers of quadrupole moments separated by triangular layers of polarized magnetic moments,
which is reminiscent of the so-called kagom\'{e} ice (KI) state of spin ice (SI) materials~\cite{MatsuhiraJPCM2002,HiroiJPSJ2002,HigashinakaPRB2003,SakakibaraPRL2003,TabataPRL2006,H.TakatsuJPSJ2013,H.OtsukaPRB2014}.

Therefore, it is intriguing to examine thermodynamic properties of the quadrupolar state in Tb$_{2+x}$Ti$_{2-x}$O$_{7+y}$ under magnetic field. 
For this purpose, we studied specific heat ($C_P$) and the entropy ($S$) change of a sample showing the LRO.
Here we focus on experiments in the [111] and [001] field directions and
show the $T$-- and $H$--dependence of $S$ and 
$\varDelta S [=S(0.55{\rm K},H) - S(0.15{\rm K},H)]$.
We found that 
the plateau-like behavior of $\varDelta S$ appears 
in fields around 0.5~T only for $H||[111]$, while the change is soon suppressed for $H||[001]$.
This plateau state is attributed to the change in states or the formation of the quadrupole order 
on the kagom\'{e} layers perpendicular to the magnetic field.

\section{Experimental}
Single crystals of Tb$_{2+x}$Ti$_{2-x}$O$_{7+y}$ were grown by a floating zone method~\cite{M.WakitaJPCS2016}.
We used the crystal with $x = 0.005$. 
Specific heat $C_P$ was measured by a quasi-adiabatic heat-pulse or thermal relaxation method.
In a temperature range below 2~K, we used a dilution, $^3$He, and an adiabatic demagnetization refrigerators,
while above 2~K we used a Quantum Design PPMS system.
The in-field data presented here were obtained using 
a vector magnet system where an accuracy of the field direction to the sample is below $1^\circ$.
In order to reduce the demagnetization effect,
we used a plate-like crystal
along the $\langle110\rangle$ plane which includes the [111], [110], and [001] axes.
The sample is approximately $0.7\times0.9\times0.1$~mm$^3$ which is 0.35~mg in weight. 
Since the demagnetization factor for the [111] and [001] directions is small enough ($N\simeq0.09$),
demagnetization corrections were not performed in the present study.

\section{Results and Discussion}
\begin{figure}
\begin{center}
\includegraphics[width=0.85\textwidth,clip]{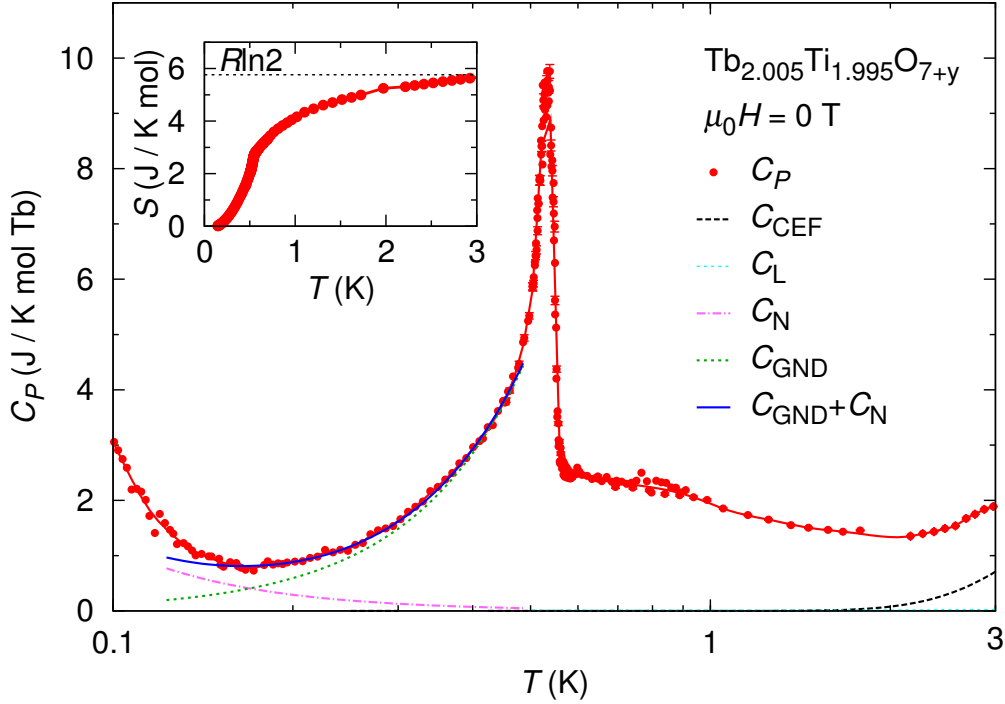}
\end{center}
\caption{
Temperature dependence of the specific heat $C_P$ of single-crystalline
Tb$_{2+x}$Ti$_{2-x}$O$_{7+y}$ with $x=0.005$ at zero field.
The red filled circles represent the measured $C_P$ data from 3 to 0.1~K.
$C_{\rm CEF}$, $C_{\rm L}$, $C_{\rm N}$, and $C_{\rm GND}$ represent 
the calculated or fitted results of the contribution from excited CEF
to the specific heat, the lattice specific heat, the nuclear specific heat, and 
the specific heat of the low-$T$ excitations of the phase transition, respectively.
Inset shows the temperature dependence of the entropy evaluated from
the integration of $(C_P - C_{\rm CEF} - C_{\rm L} -C_{\rm N})/T$ from 0.15~K.
}
\label{fig.1}
\end{figure}
Temperature dependence of $C_P$ at zero field is shown in Fig.~\ref{fig.1}.
The specific heat $C_P$ exhibits a minimum around 2~K and slightly increases on cooling toward $T_{\rm c}= 0.53$~K,
implying a Schottky-type anomaly characterized by SI correlations~\cite{HarrisPRL1998}.
A clear peak at $T_{\rm c}$ results in the phase transition to the quadrupolar state~\cite{H.TakatsuPRL2016}.
These behaviors are compatible with the experimental results of the polycrystalline sample of $x=0.005$~\cite{TaniguchiPRB2013}.

In order to analyze the low-$T$ behavior of the specific heat of 
the ground state doublet ($C_{\rm GND}$) and its entropy change, it is important to estimate 
other contributions and subtract those from the measured $C_P$ data. 
The specific heat of insulating Tb$_{2+x}$Ti$_{2-x}$O$_{7+y}$ 
is represented as $C_P = C_{\rm GND} + C_{\rm CEF} + C_{\rm L} + C_{\rm N}$.
Here $C_{\rm CEF}$ is a contribution attributed to higher-energy crystal electric field (CEF) states.
This contribution was calculated by taking the CEF scheme obtained by Ref.~\cite{MirebeauPRB2007} and 
assuming the Schottky specific heat. 
It is slightly visible at temperatures above 2~K and becomes negligible below 1~K.
$C_{\rm L}$ is the lattice specific heat estimated in the same way as 
described in Ref.~\cite{GingrasPRB2000}.
As reference, this is negligibly small at low temperatures (viz. below 3~K)~\cite{GingrasPRB2000,M.RuminyPRB2016-2}.
$C_{\rm N}$ is the nuclear specific heat showing the Schottky anomaly.
It is finite in a LRO state with the finite magnetic-dipole or electric-quadrupole hyperfine field from 4$f$ moment~\cite{Elliott1953,Bleaney1961}
or in external magnetic field~\cite{E.S.R.Gopal}, because of
the nuclear level splitting of the nuclear spin $I=3/2$ of $^{159}$Tb with the natural abundance 100\%,
which mostly contributes to $C_{\rm N}$.
For this estimation, we fitted the low-$T$ part in the range between 0.1 and 0.4~K using the relation of 
$C_P \simeq C_{\rm GND} + C_{\rm N} = AT^{n} + B/T^{2}$, where $C_{\rm GND}$ is assumed to be the power law
temperature dependence with respect to low-$T$ excitations for the phase transition,
and $C_{\rm N}$ is tentatively used as the term proportional to $1/T^2$: 
$A$ and $B$ are the coefficients of these $T$ dependences.
One of the fitting results is shown in Fig.~1. 
The low-$T$ behavior down to 0.15~K can be fitted by this simple relation.
However, it is not applicable for the data below 0.15 K.
Although we extended to fit the data using higher order terms of $C_{\rm N}$
obtained by both considerations of magnetic dipole hyperfine coupling 
and electric nuclei-quadrupole coupling~\cite{LounasmaaPR1962},
the fitting was not improved.
This may reflect low energy fluctuations or other anomalous contributions
such as photon-like excitations and the proximity of quantum criticality~\cite{HermelePRB2004,TaniguchiPRB2013}.
It is also considered that a coupling between low energy fluctuations of 
$4f$ moments and nuclear spins (and also quadrupole moments) of the Tb nuclei
may give rise to the anomalous enhancement of $C_P$. 
Indeed, the energy scale of 0.1~K corresponds to 10~$\mu$eV
and then the inelastic neutron spectrum previously observed around $E=0$~\cite{TaniguchiPRB2013}
could be ascribable to the possible existence of such low energy excitations, 
although higher resolution experiments are needed to clarify 
this point.
Otherwise, the anomalous enhancement of $C_P$ might be related to 
the lowering of the thermal conductivity of the sample on cooling,
which may cause temperature gradient inside the sample and overestimation of the $C_P$ value,
although the small and thin crystal was used for the experiments. 
Note that such a large enhancement of $C_P$ has been also observed in 
a QSL of the metallic
pyrochlore Pr$_{2}$Ir$_{2}$O$_7$~\cite{TokiwaNatMat2014}.
More precise and careful measurements are required 
for further understanding.
The fitting yields $A = 24(2)$, $n = 2.4 (1)$, and $B = 0.014(2)$.
Since the coefficient $B$ is written to be $B = R(1.25 \alpha'^2 + P^2)$
using the magnetic hyperfine constant $\alpha'$, quadrupole coupling constant $P$, and gas constant $R$~\cite{LounasmaaPR1962},
it is roughly estimated that $\alpha' \sim 0.037$~K when $P = 0$ 
or $P \sim 0.041$~K when $\alpha' = 0$ (or then $\alpha'$ and $P$ are expected within half values of those). 
These values are the same order of the values for systems including Tb nuclei 
and the theoretical expectation for the case of the Tb metal~\cite{J.KondoJPSJ1961,BleaneyJAP1963}. 
We thus considered that the estimation of $C_N$ is approximately reasonable and 
intended to focus on the data above 0.15~K at present for the evaluation of the entropy,
which was calculated 
from the integration of $\varDelta C/T = (C_P - C_{\rm CEF} - C_{\rm L} -C_{\rm N})/T$.
The corresponding entropy change in temperature is shown in the inset of Fig.~\ref{fig.1}.
%

\begin{figure}
\begin{center}
\includegraphics[width=0.85\textwidth,clip]{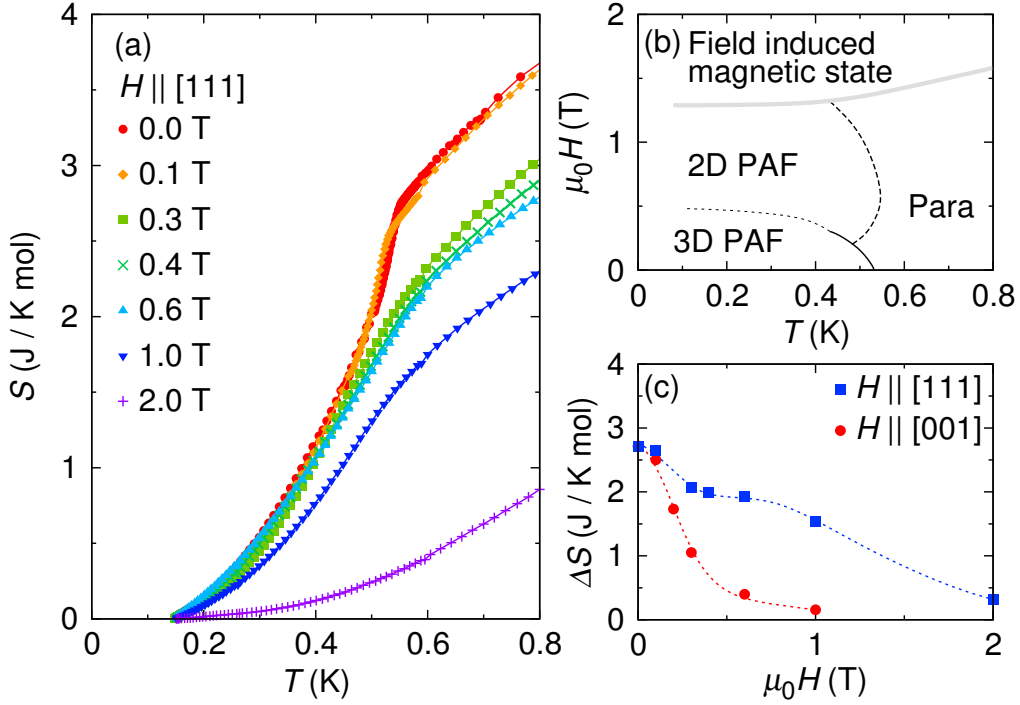}
\end{center}
\caption{
(a)Temperature dependence of the entropy for several magnetic fields for $H||[111]$.
(b)$H$--$T$ phase diagram for $H||[111]$~\cite{H.TakatsuPRL2016}. 
The black solid line up to 0.4 T indicates the phase boundary between the 3D quadrupolar 
state of the planar antiferropseudospin (PAF), 3D PAF, and the paramagnetic paraquadrupole state.
The black dashed line is the phase boundary between the 2D quadrupolar state, 2D PAF, 
and the paramagnetic paraquadrupole state.
The black dotted line represents the line of the crossover or phase transition 
between the 3D and 2D PAF states. The gray solid line is 
the putative phase boundary between the 2D PAF state and the field-induced magnetic state, 
which is estimated from a kink of the magnetization.
(c)Field dependence of the entropy change $\varDelta S [=S(0.55{\rm K},H) - S(0.15{\rm K},H)]$
for $H||[111]$ and $H||[001]$. The lines are guides to the eyes.
}
\label{fig.2}
\end{figure}

It is remarkable that the entropy in zero field at $T_{\rm c}$ is 2.7~J~K$^{-1}$mol$^{-1}$.
This value is about 50\% of $R\ln2$ expected for a non-Kramers doublet of the Tb ions. 
Even when we consider the low-$T$ contribution below 0.15~K,
it doesn't reach $R\ln2$.
Instead, 
the entropy release persists to higher temperatures and is saturated around 3~K, as seen in the inset of Fig.~\ref{fig.1}.
This behavior is similar to that of previous experiments for a sample showing LRO~\cite{HamaguchiPRB2004}.
These results imply that strong fluctuations still persist to temperatures higher than $T_{\rm c}$.
It is also suggested that a QSL-like state or the proximity of it could realize 
at temperatures between near $T_{\rm c}$ and $\sim1$~K,
where $C_{P}$ slightly increases like a Schottky anomaly signified by SI correlations
and the value is also large enough ($\sim2$~~J~K$^{-1}$mol-Tb$^{-1}$): these imply finite excitation
density of states or large magnetic fluctuations of quantum or thermally-excited monopoles,
although $S$ does not exhibit the plateau of the residual entropy, 
which is masked or must be released by the sharp peak of $C_P$ at $T_{\rm c}$.
%

Figure~\ref{fig.2}(a) shows the temperature dependence of the field variation
of the entropy, $S(T,H)$, for $H||[111]$.
For these estimations, we used the same method described above,
assuming that $C_{\rm CEF}$ and $C_{\rm L}$ in fields
are also negligibly small at low temperatures below 1~K. 
Interestingly, 
it is found that $S(T,H)$ exhibits quantitatively the same behavior in fields around 0.5~T,
while it decreases (or increases) with increasing (or decreasing) magnetic fields in a temperature range up to 0.8~K. 
These behaviors reflect the properties of the $H$--$T$ phase diagram for $H||[111]$ [Fig.~\ref{fig.2}(b)],
where the 3D quadrupolar state is considered to be replaced by the 2D quadrupolar state, and then by the magnetic state
at low temperatures. 
In fact, as plotted in Fig.~\ref{fig.2}(c),
the field dependence of the entropy change around zero-field $T_{\rm c}$, $\varDelta S$, 
exhibits a plateau-like behavior around 0.5~T for $H||[111]$.
This decreases rapidly for $H||[001]$ 
reflecting the relatively small field that induces
the magnetic state above 0.3~T~\cite{H.TakatsuJPCS2016}.
It is considered that at the intermediate field ($\sim0.5$~T) for $H||[111]$,
the phase transition occurs only due to the change of the state in each kagom\'{e}-lattice layer, 
which could only contribute to the change in entropy.
Therefore, this situation may lead to almost the same temperature dependence of $S(T,H)$ around 0.5~T and the plateau state of $\varDelta S$.

Now we find that the difference of $\varDelta S$ between at 0 and around 0.5~T for $H||[111]$ 
is about 0.7~J~K$^{-1}$mol$^{-1}$. Then, a question is what does this value mean?
An interesting scenario is that, 
if the states in 0 and about 0.5~T around 1~K  are similar to the classical SI and KI states and
the residual entropies of these states are released by the phase transition and then zero as expected from 
the quantum model~\cite{Y.KatoPRL2015}, 
the difference of $\varDelta S$ between at 0 and around 0.5~T may possibly correspond to 
the difference between the SI residual entropy (1.68~J~K$^{-1}$mol$^{-1}$)~\cite{RamirezNature1999}
and the KI residual entropy (0.67~J~K$^{-1}$mol$^{-1}$)~\cite{UdagawaJPSJ2002,MoessnerPRB2003}; i.e., $\sim1$~J~K$^{-1}$mol$^{-1}$.
In fact, the value of 0.7~J~K$^{-1}$mol$^{-1}$ is close to this value. 
It is considered that this entropy will be released at temperatures higher than 0.8~K.
For this context, 
further detailed experiments at higher temperatures under magnetic field are interesting future subjects.

\section{Conclusion}
In conclusion, 
we reported thermodynamic properties of the quadrupole-order sample of Tb$_{2+x}$Ti$_{2-x}$O$_{7+y}$ with $x=0.005$.
We observed that the specific heat shows a minimum around 2~K and slightly increases on cooling,
which is similar to 
a Schottky-type anomaly for canonical spin ices.
A clear peak at $T_{\rm c}=0.53$~K results in the phase transition to the quadrupolar state.
The entropy change at $T_{\rm c}$ is 2.7~J~K$^{-1}$mol$^{-1}$, which is smaller than $R\ln2$ and suggests 
the strong fluctuations that remains at higher temperatures than $T_{\rm c}$.
The field dependence of the entropy change 
around 0.5~K and in 0.5~T for $H||[111]$ exhibits a plateau.
This characteristic feature probably
reflects the formation of the two-dimensional quadrupolar state by the [111] magnetic field.
%

\ack 
We thank S. Onoda, Y. Kato, R. Higashinaka, M. Wakita, and H. Kageyama for useful discussions
and their assistance. 
This work was supported by JSPS KAKENHI grant numbers 25400345 and 26400336.
One of the specific heat measurements was performed using facilities of ISSP, University of Tokyo.
One of the authors would like to acknowledge the support from the Motizuki Fund of Yukawa Memorial Foundation.
\section*{References}
\bibliography{20170424_TTO_HT_HFM_proceedings.bbl}

\providecommand{\newblock}{}
\begin{thebibliography}{10}
\expandafter\ifx\csname url\endcsname\relax
  \def\url#1{{\tt #1}}\fi
\expandafter\ifx\csname urlprefix\endcsname\relax\def\urlprefix{URL }\fi
\providecommand{\eprint}[2][]{\url{#2}}

\bibitem{C.Lacroix}
Lacroix C, Mendels P and Mila F (eds) 2011 {\em Introduction to Frustrated
  Magnetism\/} (Berlin, Heidelberg: Springer)

\bibitem{GingrasRPP2014}
Gingras M~J~P and McClarty P~A 2014 {\em Rep. Prog. Phys.\/} {\bf 77} 056501

\bibitem{GardnerPRL1999}
Gardner J~S, Dunsiger S~R, Gaulin B~D, Gingras M~J~P, Greedan J~E, Kiefl R~F,
  Lumsden M~D, MacFarlane W~A, Raju N~P, Sonier J~E, Swainson I and Tun Z 1999
  {\em Phys. Rev. Lett.\/} {\bf 82} 1012

\bibitem{GardnerPRB2003}
Gardner J~S, Keren A, Ehlers G, Stock C, Segal E, Roper J~M, Fak B, Stone M~B,
  Hammar P~R, Reich D~H and Gaulin B~D 2003 {\em Phys. Rev. B\/} {\bf 68}
  180401

\bibitem{TaniguchiPRB2013}
Taniguchi T, Kadowaki H, Takatsu H, Fak B, Ollivier J, Yamazaki T, Sato T~J,
  Yoshizawa H, Shimura Y, Sakakibara T, Hong T, Goto K, Yaraskavitch L~R and
  Kycia J~B 2013 {\em Phys. Rev. B\/} {\bf 87} 060408(R)

\bibitem{M.WakitaJPCS2016}
Wakita M, Taniguchi T, Edamoto H, Takatsu H and Kadowaki H 2016 {\em J. Phys.:
  Conf. Ser.\/} {\bf 683} 012023

\bibitem{GingrasPRB2000}
Gingras M~J~P, den Hertog B~C, Faucher M, Gardner J~S, Dunsiger S~R, Chang L~J,
  Gaulin B~D, Raju N~P and Greedan J~E 2000 {\em Phys. Rev. B\/} {\bf 62} 6496

\bibitem{KaoPRB2003}
Kao Y~J, Enjalran M, Maestro A~D, Molavian H~R and Gingras M~J~P 2003 {\em
  Phys. Rev. B\/} {\bf 68} 172407

\bibitem{MolavianPRL2007}
Molavian H~R, Gingras M~J~P and Canals B 2007 {\em Phys. Rev. Lett.\/} {\bf 98}
  157204

\bibitem{BonvillePRB2011}
Bonville P, Mirebeau I, Gukasov A, Petit S and Robert J 2011 {\em Phys. Rev.
  B\/} {\bf 84} 184409

\bibitem{PetitEPJWC2015}
Petit S, Guitteny S, Robert J, Bonville P, Decorse C, Ollivier J, Mutka H and
  Mirebeau I 2015 {\em EPJ Web Conf.\/} {\bf 83} 03012

\bibitem{GuittenyPRB2015}
Guitteny S, Mirebeau I, de~R$\rm{\acute{e}}$otier P~D, Colin C~V, Bonville P,
  Porcher F, Grenier B, Decorse C and Petit S 2015 {\em Phys. Rev. B\/} {\bf
  92} 144412

\bibitem{FritschPRB2013}
Fritsch K, Ross K~A, Qiu Y, Copley J~R~D, Guidi T, Bewley R~I, Dabkowska H~A
  and Gaulin B~D 2013 {\em Phys. Rev. B\/} {\bf 87} 094410

\bibitem{FritschPRB2014}
Fritsch K, Kermarrec E, Ross K~A, Qiu Y, Copley J~R~D, Pomaranski D, Kycia J~B,
  Dabkowska H~A and Gaulin B~D 2014 {\em Phys. Rev. B\/} {\bf 90} 014429

\bibitem{E.KermarrecPRB2015}
Kermarrec E, Maharaj D~D, Gaudet J, Fritsch K, Pomaranski D, Kycia J~B, Qiu Y,
  Copley J~R~D, Couchman M, Morningstar A, Dabkowska H~A and Gaulin B~D 2015
  {\em Phys. Rev. B\/} {\bf 92} 245114

\bibitem{FennellPRL2014}
Fennell T, Kenzelmann M, Roessli B, Mutka H, Ollivier J, Ruminy M, Stuhr U,
  Zaharko O, Bovo L, Cervellino A, Haas M~K and Cava R~J 2014 {\em Phys. Rev.
  Lett.\/} {\bf 112} 017203

\bibitem{M.RuminyPRB2016-1}
Ruminy M, Bovo L, Pomjakushina E, Haas M~K, Stuhr U, Cervellino A, Cava R~J,
  Kenzelmann M and Fennell T 2016 {\em Phys. Rev. B\/} {\bf 93} 144407

\bibitem{M.Ruminy2016}
Ruminy M, Groitl F, Keller T and Fennell T {}arXiv:1607.07688

\bibitem{KadowakiSPIN2015}
Kadowaki H, Takatsu H, Taniguchi T, F{\aa}k B and Ollivier J 2015 {\em SPIN\/}
  {\bf 5} 1540003

\bibitem{H.TakatsuPRL2016}
Takatsu H, Onoda S, Kittaka S, Kasahara A, Kono Y, Sakakibara T, Kato Y,
  F{\aa}k B, Lynn J~O~J~W, Taniguchi T, Wakita M and Kadowaki H 2016 {\em Phys.
  Rev. Lett.\/} {\bf 116} 217201

\bibitem{S.Onoda2010PRL}
Onoda S and Tanaka Y 2010 {\em Phys. Rev. Lett.\/} {\bf 105} 047201

\bibitem{S.Onoda2011PRB}
Onoda S and Tanaka Y 2011 {\em Phys. Rev. B\/} {\bf 83} 094411

\bibitem{MatsuhiraJPCM2002}
Matsuhira K, Hiroi Z, Tayama T, Takagi S and Sakakibara T 2002 {\em J. Phys.:
  Condens. Matter\/} {\bf 14} L559

\bibitem{HiroiJPSJ2002}
Hiroi Z, Matsuhira K, Takagi S, Tayama T and Sakakibara T 2002 {\em J. Phys.
  Soc. Jpn.\/} {\bf 72} 411

\bibitem{HigashinakaPRB2003}
Higashinaka R, Fukazawa H and Maeno Y 2003 {\em Phys. Rev. B\/} {\bf 68} 014415

\bibitem{SakakibaraPRL2003}
Sakakibara T, Tayama T, Hiroi Z, Matsuhira K and Takagi S 2003 {\em Phys. Rev.
  Lett.\/} {\bf 90} 207205

\bibitem{TabataPRL2006}
Tabata Y, Kadowaki H, Matsuhira K, Hiroi Z, Aso N, Ressouche E and F{\aa}k B
  2006 {\em Phys. Rev. Lett.\/} {\bf 97} 257205

\bibitem{H.TakatsuJPSJ2013}
Takatsu H, Goto K, Otsuka H, Higashinaka R, Matsubayashi K, Uwatoko Y and
  Kadowaki H 2013 {\em J. Phys. Soc. Jpn.\/} {\bf 82} 073707

\bibitem{H.OtsukaPRB2014}
Otsuka H, Takatsu H, Goto K and Kadowaki H 2014 {\em Phys. Rev. B\/} {\bf 90}
  144428

\bibitem{HarrisPRL1998}
Harris M~J, Bramwell S~T, Holdsworth P~C~W and Champion J~D~M 1998 {\em Phys.
  Rev. Lett.\/} {\bf 81} 4496

\bibitem{MirebeauPRB2007}
Mirebeau I, Bonville P and Hennion M 2007 {\em Phys. Rev. B\/} {\bf 76} 184436

\bibitem{M.RuminyPRB2016-2}
Ruminy M, Valdez M~N, Wehinger B, Bosak A, Adroja D~T, Stuhr U, Iida K,
  Kamazawa K, Pomjakushina E, Prabakharan D, Haas M~K, Bovo L, Sheptyakov D,
  Cervellino A, Cava R~J, Kenzelmann M, Spaldin N~A and Fennell T 2016 {\em
  Phys. Rev. B\/} {\bf 93} 214308

\bibitem{Elliott1953}
Elliott R~J and Stevens K~W~H 1953 {\em Proc. Roy. Soc. A\/} {\bf 218} 553

\bibitem{Bleaney1961}
Bleaney B and Hill R~W 1961 {\em Proc. Roy. Soc.\/} {\bf 78} 313

\bibitem{E.S.R.Gopal}
Gopal E~S~R 1966 {\em Specific Heats at Low Temperatures\/} (New York: ~Plenum
  Press)

\bibitem{LounasmaaPR1962}
Lounasmaa O~V and Roach P~R 1962 {\em Phys. Rev.\/} {\bf 128} 622

\bibitem{HermelePRB2004}
Hermele M, Fisher M~P~A and Balents L 2004 {\em Phys. Rev. B\/} {\bf 69} 064404

\bibitem{TokiwaNatMat2014}
Tokiwa Y, Ishikawa J~J, Nakatsuji S and Gegenwart P 2014 {\em Nature Mater.\/}
  {\bf 13} 356

\bibitem{J.KondoJPSJ1961}
Kondo J 1961 {\em J. Phys. Soc. Jpn.\/} {\bf 16} 1690

\bibitem{BleaneyJAP1963}
Bleaney B 1963 {\em J. Appl. Phys.\/} {\bf 16} 1690

\bibitem{HamaguchiPRB2004}
Hamaguchi N, Matsushita T, Wada N, Yasui Y and Sato M 2004 {\em Phys. Rev. B\/}
  {\bf 69} 132413

\bibitem{H.TakatsuJPCS2016}
Takatsu H, Taniguchi T, Kittaka S, Sakakibara T and Kadowaki H 2016 {\em J.
  Phys.: Conf. Ser.\/} {\bf 683} 012023

\bibitem{Y.KatoPRL2015}
Kato Y and Onoda S 2015 {\em Phys. Rev. Lett.\/} {\bf 115} 077202

\bibitem{RamirezNature1999}
Ramirez A~P, Hayashi A, Cava R~J, Siddharthan R and Shastry B~S 1999 {\em
  Nature\/} {\bf 399} 333

\bibitem{UdagawaJPSJ2002}
Udagawa M, Ogata M and Hiroi Z 2002 {\em J. Phys. Soc. Jpn.\/} {\bf 71} 2365

\bibitem{MoessnerPRB2003}
Moessner R and Sondhi S~L 2003 {\em Phys. Rev. B\/} {\bf 68} 184512

\end{thebibliography}

\end{document}